\documentclass[10pt]{iopart}
\usepackage{iopams}
\usepackage[utf8]{inputenc}
\usepackage{color}
\usepackage{overpic}
\usepackage{soul}
\usepackage{cite}

\usepackage{graphicx}
\usepackage[numbers]{natbib}

\usepackage{float}
\usepackage{graphicx}
\graphicspath{{./figures/}}
\usepackage[export]{adjustbox}

\usepackage{pgfplots}
\pgfplotsset{compat=1.15}
\usepackage{pgfplotstable} 
\usepgfplotslibrary{external} 
\usepackage[breaklinks=true]{hyperref}
\hypersetup{
  unicode=false,            pdftoolbar=true,          pdfmenubar=true,          pdffitwindow=false,       pdfstartview={FitH},      pdfsubject={Subject},     pdfcreator={Creator},     pdfproducer={Producer},   pdfkeywords={keyword1} {key2} {key3},   pdfnewwindow=true,        colorlinks=true,          linkcolor=blue,           citecolor=blue,           filecolor=blue,           urlcolor=blue           }

\begin{document}

\title[O.~Vallhagen et al, SPI simulations with plasmoid drifts in AUG and ITER]{Simulation of Shattered Pellet Injections with Plasmoid Drifts in ASDEX Upgrade and ITER
  }
\author{O.~Vallhagen$^1$, L.~Antonsson$^1$, P.~Halldestam$^{2}$, G.~Papp$^2$, P.~Heinrich$^2$, A.~Patel$^2$, M.~Hoppe$^3$, L.~Votta$^3$, the ASDEX Upgrade Team$^a$, the EUROfusion Tokamak Exploitation Team$^b$}
\address{$^1$ Department of Physics, Chalmers University of Technology,
  SE-41296 Gothenburg, Sweden}
\address{$^2$ Max Planck Institute for Plasma Physics, Garching, Germany}
\address{$^3$Department of Electrical Engineering, KTH Royal Institute of Technology, SE-11428 Stockholm, Sweden}
\address{$^a$See author list of H. Zohm \emph{et al}, NF 2024} 
\address{$^b$See author list of E. Joffrin \emph{et al}, NF 2024}
    \ead{vaoskar@chalmers.se}
\begin{abstract}
Pellet injection is an important means to fuel and control discharges and mitigate disruptions in reactor-scale fusion devices. To accurately assess the efficiency of these applications, it is necessary to account for the drift of the ablated material toward the low-field side. In this study, we have implemented a semi-analytical model for ablation cloud drifts in the numerical disruption modelling tool DREAM. We show that this model is capable of reproducing the density evolution in shattered pellet injection (SPI) experiments in ASDEX Upgrade, for model parameters within the expected range. The model is then used to investigate the prospects for disruption mitigation by staggered SPIs in $15\,\rm MA$ DT H-mode ITER scenarios. We find that the drifts may decrease the assimilation of pure deuterium SPIs by about an order of magnitude, which may be important to consider when designing the disruption mitigation scheme in ITER. The ITER scenarios studied here generally result in similar multi-MA runaway electron (RE) currents, regardless of the drift assumptions, but the effect of the drift is larger in situations with a fast and early thermal quench. The RE current may also be more strongly affected by the drift losses when accounting for RE losses caused by the vertical plasma motion.
\end{abstract}

\ioptwocol

\section{Introduction}
\label{sec:introduction}
Pellet injection is an important means to fuel and control discharges and mitigate disruptions in
reactor-scale fusion devices. The disruption mitigation system in ITER is based on Shattered Pellet Injection (SPI), where pellets consisting of a mixture of hydrogen~(H) and neon~(Ne) will be shattered against a bend at the end of the guide tube before entering the plasma \cite{LehnenITER}. Shattering the pellet enables optimising the assimilation efficiency in order to quickly reach the densities required. This is particularly important for disruption mitigation, as the density increase must be sufficient to safely dissipate the thermal and magnetic energy content of the plasma uniformly through radiation, and increase the frictional drag sufficiently to reduce runaway electron (RE) formation. 

Since the SPI technique was first demonstrated at the DIII-D tokamak in 2010 \cite{Commaux2010}, similar systems have been installed at several devices, including DIII-D \cite{Meitner2017}, ASDEX Upgrade (AUG) \cite{Dibon2023, Heinrich_2024_SPI_Lab, Heinrich_2025_frad, Heinrich_2025_PhD}, KSTAR~\cite{Park2020} and JET~\cite{Baylor2021, WilsonSOFT}. These systems have recently been used extensively to conduct experiments in support of the ITER disruption mitigation system \cite{Herfindal2019, Papp2021, Jachmich2022, Baylor2021}. While promising results have been demonstrated, both regarding the primary thermal~\cite{Heinrich_2025_frad, Schwarz_2023} and magnetic load mitigation~\cite{Schwarz_2023} and RE avoidance as well as the termination of an already formed RE beam \cite{ReuxBenign, Sheikh_2024}, the scalability to reactor-scale devices remains an open question.

To assess the scaling of the SPI technique as a method for disruption mitigation it is crucial to have an accurate modelling capability. Several studies of integrated full-scenario modelling of SPI injections in ITER have previously been made with e.g. the DREAM \cite{Vallhagen2022Effect, Vallhagen_2024}, JOREK \cite{Hu2021Radiation} and INDEX \cite{AkinobuITPA, Matsuyama_2022} codes. These studies used an analytical model for the pellet ablation based on the Neutral Gas Shielding (NGS) model \cite{Parks_Thurnbull, Parks_TSDW}. A particle-based pellet ablation model with an explicit tracking of the solid-gas interface at the pellet surface has been implemented in the NIMROD code \cite{Kim_2019}. 

The SPI dynamics are, however, complicated by a $\nabla B$-induced drift of the recently ionized ablation cloud towards the Low Field Side (LFS) \cite{lang_high-efficiency_1997,Parks2000Radial, Pegourie2006Homogenization, Vallhagen2023Drift}. This phenomenon arises due to the charge separation caused by the excess $\nabla B$-current inside the overpressurized ionized ablation cloud, which in turn gives rise to an $E\times B$-drift. The dynamics are further complicated by the formation of striations along the magnetic field lines in the cloud of drifting material, to some extent separating it into discrete plasmoids. Several mechanisms have been proposed to play a role in forming these plasmoids, such as the limited energy reservoir available on resonant magnetic flux surfaces \cite{Kaufmann_1986, Pegourie_1989}, the periodic passing of the ablation cloud front by the pellet \cite{Durst_1990, Muller}, and repeated Rayleigh-Taylor instabilities displacing the ablation cloud from the pellet \cite{ParksStriations}. The drifting of the ablating material is therefore commonly referred to as \emph{plasmoid drift}. This effect has been either ignored or accounted for with \emph{ad hoc} methods in the mentioned studies. Although the drift is intrinsically included in 3D JOREK and NIMROD simulations, it has so far not been feasible to resolve realistically small ionized ablation clouds, which would be required to accurately account for the drift \cite{Kong_2025}.

To accurately account for the variety of length scales spanned as the plasmoid expands along, and drifts across, the field lines, it is instead advantageous to use Lagrangian particle codes to simulate the plasmoid drift, as done by e.g.~\cite{Samulyak2021Lagrangian}. Such advanced simulations offer an accurate description of the ablation and transport of the pellet material, but have so far not been used in integrated modelling of full SPI scenarios. The computational complexity can be reduced considerably by using a single-cell Lagrangian model such as that of \cite{Pegourie2006Homogenization, Akinobu_2022}, which was used to succesfully replicate the density rise following fueling pellet injections in the Tore Supra tokamak. 

However, when performing SPI simulations including $\sim 1000$ shards and scanning a large parameter space, it is useful to be able to have an analytical model for estimating the drift, which can be implemented in an integrated framework without substantially increasing the computational cost. Several such models have been developed, adding different mechanisms involved in regulating the drift. An early model for the drift of a plasmoid with a homogeneous cross section was derived in \cite{Parks2000Radial}. This model accounted for the $\nabla B$-induced acceleration, the opposing polarisation current counteracting the charge separation, and the pressure equilibration with the background plasma. In addition, the drag due to the energy dissipation associated with the excitation of Alfv\'en waves propagating from the end caps of the plasmoid was also taken into account. A large aspect ratio and a completely toroidal magnetic field was assumed, i.e. neglecting the rotational transform. 

The effect of the rotational transform was added approximately in \cite{Rozhansky2004}. Here it was recognized that when the plasmoid has expanded to cover all poloidal angles, the $\nabla B$-current between two given field lines in different parts of the cloud cancel each other, such that the cloud acceleration vanishes. The time it takes for the cloud to expand one poloidal turn was therefore used as an estimate for the duration of the drift. The transient nature of the effect of the rotational transform was, however, not accounted for. Moreover, the drag due to the Ohmic currents flowing along field lines connecting the toroidally opposite end caps of the cloud \cite{Pegourie2006Homogenization} was also not considered.

These effects were later included in \cite{Vallhagen2023Drift}. This model considered the ablated material to take the form of a continuous flowing cloud with a slab geometry, and the flow velocity was found by rewriting the MHD force balance equation into a current balance equation. The poloidal electric field, and hence the drift velocity, is then set by the balance of the $\nabla B$-drift giving rise to the charge separation, the opposing polarisation current, and the current exiting the cloud parallel to the magnetic field lines. The resistivity for the Ohmic part of the parallel current was calculated using a statistical model for the number of turns a field line circles around the torus before it reconnects to the cloud, as detailed in \cite{Vallhagen2023Drift}. 
This model was able to reproduce reasonably well an estimate of the drift from interpretive modelling of SPI experiments at JET \cite{Kong_interpretive}, as well as simulations of the drift in ITER \cite{Akinobu_2022}. The latter indicated that the drift could potentially have a major impact on the assimilation of pure deuterium~(D) pellets in reactor-grade tokamaks. However, it remains to study the effect of the drift in an integrated model for SPI-mitigated disruptions, which is the subject of the present paper.

In this paper, we have implemented the drift model derived in \cite{Vallhagen2023Drift} in the 1D flux surface averaged disruption modelling tool DREAM. This implementation is first used to simulate SPI experiments at the AUG tokamak, previously studied with DREAM in \cite{Halldestam2025}, and compare the results with and without the drift with the experimental data. The model is then applied to ITER scenarios, previously studied without the drift in \cite{Vallhagen_2024}, assessing the impact of the drift in different scenarios and potential implications for the ITER disruption mitigation system. The model is described in section~\ref{sec:model}, the general simulation settings are presented in section~\ref{sec:settings}, and the simulation results are presented in section~\ref{sec:results}. The results are then followed by a discussion in section~\ref{sec:discussion} before the paper is concluded in section~\ref{sec:conclusion}.

\section{Disruption model}
\label{sec:model}

\subsection{SPI model}
\label{sec:SPI}
The SPI model implemented in DREAM is described in detail in \cite{Vallhagen2022Effect}, and the relevant features are summarised below. The model takes as an input the pellet size and composition, the coordinates of the shatter point and the speed of each shard. In this paper, the velocity distribution of the shards is assumed to be either Gaussian or uniform in speed, and uniform in divergence angle, as further specified in section \ref{sec:settings}. Their sizes follow the distribution derived in \cite{Parks2016Modeling}. 

The characteristic shard size can either be specified by directly providing the desired number of shards, or be calculated using the model from \cite{Gebhart2020} based on the pellet injection speed, pellet composition and the angle of the bend at the end of the shattering tube. In the former case, we sample a fixed number of shards and rescale their sizes to match the full pellet volume, in order to be consistent with the specified model parameters. In the latter case, shards are sampled iteratively until the next shard would exceed the full pellet volume. Thus, the number of shards can differ somewhat between different realisations.

Once the pellet is shattered and the shards enter the plasma, their ablation is calculated using the NGS model presented in \cite{Parks_TSDW}, and the location of deposition following the plasmoid drift (assumed to be instantaneous) is calculated using the model of \cite{Vallhagen2023Drift}. This model assumes that the drifting cloud of ablated material has the form of a slab of constant thickness, expanding parallel to the field lines, with homogeneous plasma parameters in the poloidal and toroidal directions.
The model requires specifying the temperature $T_0$ of the cloud in the close vicinity of the pellet, its representative temperature $T$ during the drift motion, the average charge state $\langle Z_i \rangle$ for every ion species $i$ included in the pellet, and the half-width $\Delta y$ of the drifting cloud. These values may be estimated based on experimental measurements \cite{Muller} and simulations \cite{Akinobu_2022} found in the literature. 

In this paper, we set $T_0=2\,\rm eV$, $\langle Z_\mathrm{D} \rangle=1$, $\langle Z_\mathrm{Ne} \rangle=2$, $T=5\,\rm eV$ for Ne-doped pellets and $T=30\,\rm eV$ for pure D pellets, for all scenarios considered. The value of $T_0$ is motivated by the fact that this is the temperature around which the ablation cloud becomes sufficiently ionized for the drift to start. The lower value of $T$ for the doped pellets is due to the fact that radiation losses from the Ne doping become very strong at $\sim 5\,\rm eV$, preventing the temperature from increasing further. This temperature is however high enough to essentially fully ionize all hydrogenic species, motivating the value for $\langle Z_\mathrm{D}\rangle$. The charge state distribution of Ne is complicated by the large variation in the plasmoid density and optical thickness during the drift motion, making the rate coefficients used for the background plasma (see section \ref{sec:TQ}) invalid during parts it. However, as the pellets studied in this paper have very small Ne fractions, the charge state distribution of the Ne only has a small impact on the total free electron density and average charge. We can therefore set an approximate constant value without significantly affect the results. As detailed in \ref{app:Delta_y}, and will be discussed later, the drift is very sensitive to the half-thickness $\Delta y$ of the drifting cloud. We will therefore vary this model parameter to either match experimental measurements or explore different plausible scenarios. However, direct measurements and simulations in the literature indicate that $\Delta y \sim 1\,\rm cm$ \cite{Muller, Akinobu_2022}. Similar values were also used in \cite{Guth_JPP} to reproduce measurements of the effect of asymmetric ablation on pellet trajectories.

For shards that do not travel along the horizontal midplane, the value given by equation (A4) in \cite{Vallhagen2023Drift} should be interpreted as a shift in the major radius direction for the part of the ionized ablation cloud that resides in the same poloidal plane as the pellet shard, rather than the shift in the minor radius. Although the drift dynamics is complicated by the fact that the ablation cloud expands along the field lines during the drift, within the present model, all material will be deposited on the same flux surface (determined by the location of the deposition in the poloidal plane of the pellet shard). This follows from the assumption that the electrostatic potential is uniform along the field lines inside the cloud, as it implies that the electric field, and hence the drift velocity, rotates as the cloud expands in such a way that the angle towards the flux surface normal remains constant along the field lines. If the calculated deposition location is outside the plasma, the ablated material is assumed to be lost, unless otherwise stated. However, our results indicate that for doped pellets, the ablated material may not drift freely across the scrape-off layer. In that context we therefore also consider a boundary condition where the drift stops at the plasma edge (more details are given in section \ref{sec:AUG}).

The deposited material is assumed to be instantaneously homogenized over the flux surface where it is deposited, in accordance with the flux surface averaged nature of the DREAM model. The drift and homogenisation processes typically take a few hundred microseconds \cite{Pegourie2006Homogenization, Pegourie_Picchiottino}, which is fairly short compared to the time scales of interest here. In cases where the drift model is applied, the material is deposited in the singly ionized charge state, to account for the fact that the cloud must be ionized in order for the drift to start. In cases where the drift model is not applied, the material is deposited in the neutral state. The evolution of the charge state distribution is then modelled by time dependent rate equations. The ionization and recombination rates are nominally taken from the ADAS database \cite{ADAS}. However, in cases with very large D injections, such as the ITER cases studied in section~\ref{sec:ITER}, the plasma may essentially be opaque to Lyman radiation \cite{Vallhagen2022Effect}. In such cases, we instead take the ionization and recombination coefficients for hydrogen species from the AMJUEL database\footnote{http://www.eirene.de/html/amjuel.html}, containing coefficients calculated under Lyman opaque conditions.

\subsection{Temperature evolution}
\label{sec:TQ}
When the ablated material is deposited in the plasma, it cools the plasma by dilution and radiation. The radiative cooling (including both ionization/recombination radiation and line radiation) is calculated using rate equations with coefficients taken from either the ADAS or the AMJUEL database, corresponding to the ionization and recombination rates discussed in section~\ref{sec:SPI}. In case any material is deposited in a non-neutral charge state, the coresponding ionization energy is also subtracted as an additional sink. The post-disruption temperature is essentially determined by the balance of these radiative losses with heating by the Ohmic current and collisions with REs.

In cases where the plasmoid drift is accounted for, the temperature profile is also affected by the absorption of heat flowing into the drifting plasmoid, and the subsequent re-deposition of the absorbed heat where the drift stops. We model this heat transport by introducing a sink at the location of the pellet and a corresponding source at the location of deposition:
\begin{eqnarray}
    &\left(\frac{dQ}{dt}\right)^\mathrm{heatabs} = \mathrm{e}^{-\Delta \phi/T_\mathrm{bg}}f_q(\alpha)2\pi\Delta y^2q\nonumber\\
    &\times\frac{f_\mathrm{redep}\delta(r-(\rho _\mathrm{p}+\Delta r))-\delta(r-\rho _\mathrm{p})}{V'},
\end{eqnarray}
where $\rho _\mathrm{p}$ is the flux surface label of the pellet position and $\Delta r$ is the shift in the minor radius coordinate caused by the drift. The factor $e^{-\Delta \phi /T_\mathrm{bg}}$ accounts for the reduction of the heat flux $q=n_\mathrm{e}\sqrt{2T_\mathrm{e}^3/(\pi m_\mathrm{e})}$ due to the electrostatic sheath forming at the end caps of the plasmoid, with $\Delta \phi \approx 1.8 T_\mathrm{bg}$~\cite{Parks2000Radial}. The factor $f_q(\alpha)$, with $\alpha = \lambda_\mathrm{mfp}/s$, is used to account for the heat flux attenuation due to the plasmoid, which has a shielding length $s$ on the field lines where the pellet resides ($\lambda_\mathrm{mfp}$ denotes the thermal mean free path inside the plasmoid). An expression for $f_q$ was derived in \cite{Guth_JPP}, yielding
\begin{eqnarray}
    f_q(\alpha) =& \frac{1}{\alpha^2}\left[ e^{\frac{-1}{\sqrt{\alpha}}}\left(-\frac{1}{2}\sqrt{\alpha}+\frac{1}{2}\alpha+\alpha^{3/2}+\alpha^2\right)\right.\nonumber\\
    &\left.-\frac{1}{2}\mathrm{Ei}\left(-\frac{1}{\sqrt{\alpha}}\right)\right],
\end{eqnarray}
where $\mathrm{Ei}(\cdot)$ is the exponential integral. Moreover, the factor $2\pi\Delta y^2$ is the cross section area of the two end caps of the plasmoid, and $\delta(\cdot)$ denotes a one-dimensional delta function. The factor $f_\mathrm{redep}$ is the fraction of the absorbed energy which is re-deposited along with the ablated material when the drift stops. We set this factor to $f_\mathrm{redep}=1$ for pure D pellets and $f_\mathrm{redep}=0$ for Ne doped pellets, as in the latter case the majority of the absorbed heat is expected to be radiated away. Finally, $V'=\partial V/\partial r$ is the derivative of the volume enclosed by the flux surface, labeled $r$, with respect to $r$.

It should be noted that this model is likely to underestimate the amount of heat absorbed, as it only accounts for the heat being absorbed in the vicinity of the pellet. Some heat absorption is likely to take place also during the later stages of the drift motion, especially for doped pellets. This effect is however expected to be smaller for pure hydrogenic pellets, since at some point the heat flux from the plasmoid back into the background plasma will start to balance the incoming heat flux. In either case, in order to accurately account for this heat absorption, the drift motion would have to be resolved during the simulation, a capability which is currently not included in the present model. Moreover, due to the scaling of the heat flux with the temperature, this mechanism is significantly reduced behind the front of the pellet shard plume where some cooling has already taken place.

In addition to the direct cooling by the ablated material, the injection is expected to destabilize MHD modes, leading to a thermal quench. We mimic this effect by prescribing a large global advective and/or diffusive transport of heat and ions during a few milliseconds after the onset of the disruption. The transport coefficients are either directly prescribed or, in case of the temperature, calculated assuming a Rechester-Rosenbluth form \cite{Rechester1978Electron} $D \propto (\delta B/B)^2$, with a prescribed value of the magnetic perturbation amplitude $\delta B/B$. The prescribed values determining the transport coefficients and the assumed criterion for the onset of the transport event is varied between different scenarios, as detailed in section~\ref{sec:settings}.

Finally, while the above cooling mechanisms act directly on the electrons, the ions are also cooled by collisional exchange of thermal energy with the electrons.

\subsection{Current evolution}
\label{sec:current}
The evolution of the plasma current is based on the mean field equation for the poloidal flux, including a hyperresistivity term to mimic the flattening of the current during the transport event \cite{Boozer2018}. The hyperresistivity parameter $\Lambda$ is set to achieve a current spike of the expected size, as detailed in section~\ref{sec:settings}. The nearest toroidally closed conducting surface imposes a boundary condition on the poloidal flux at a minor radius $b>a$, where $a$ is the minor radius of the plasma. The inductive coupling to this wall is determined by the resistive time scale $t_\mathrm{wall}$, which is also detailed in section~\ref{sec:settings}.

The plasma current is divided into an Ohmic part and a part carried by REs. The DREAM code currently does not include models for non-inductive current drive, and these currents are therefore incorporated into the Ohmic current density. Note, however, that once these current driving mechanisms vanish as the plasma disrupts, they will induce an electric field, driving an Ohmic current, which is expected to essentially replace the non-inductive current drive. Thus, this simplification is not expected to have a major impact on the simulation results.

In this study we use the fluid mode in DREAM (see \cite{Hoppe2021DREAM} for more details). In this model, the RE current is calculated using analytical models for the generation rate of the density of the RE population, which is assumed to stream along the field lines at the speed of light. The sources of REs include the Dreicer mechanism, for which the growth rate is calculated using a neural network trained on kinetic simulation data \cite{NN_Dreicer}, and the hot-tail mechanism, calculated using the model described in appendix C of \cite{Hoppe2021DREAM}. In the case of a DT plasma, REs are also generated by tritium decay \cite{Elongation, Martin2017Formation} and Compton scattering by $\gamma$-photons from the radioactive wall \cite{Martin2017Formation}. Finally, an existing seed of REs is amplified by the avalanche mechanism, due to large angle collisions between existing REs and bulk plasma electrons, using the model derived in \cite{Hesslow_fluid_ava}.

Once generated, REs can be lost due to collisional slowing-down when the electric field falls below the critical electric field for RE generation \cite{Hesslow_ECeff}. This is accounted for by letting the avalanche growth rate become negative at such electric fields, as detailed in \cite{Hesslow_fluid_ava}. During the transport event, REs can also be lost by diffusive transport due to the stochastization of the magnetic field lines. This transport is accounted for by either using a diffusion coefficient directly prescribed by the user, or using the model of \cite{Svensson2021Effects}. In the latter, the momentum dependent diffusion coefficient is translated to a diffusion coefficient for the total fluid RE density by integration over momentum space, using an analytical model of the distribution function assuming an avalanche-dominated RE generation. We assume a momentum dependence of the diffusion coefficient of the form $D(p) = D_0 p/(1+p^2)$, where $p$ is the magnitude of the momentum normalised to $m_e c$, with $m_e$ denoting the electron mass. For $D_0$ we use the Rechester-Rosenbluth diffusion coefficient for an electron travelling at the speed of light. This momentum dependence captures both the linear momentum dependence at non-relativistic momenta, and the decrease in the diffusion coefficient at high momenta due to orbit decorrelation effects \cite{Hauff_Jenko}.

\section{Scenarios and settings}
\label{sec:settings}

\subsection{AUG scenarios}
\label{sec:AUGsettings}

The AUG simulations presented in this paper use similar settings as in \cite{Halldestam2025}, except for the drift dynamics, which were not included there. The input profiles for the electron and ion densities and temperatures, current and magnetic field geometry are taken from discharge \#40655 (more details in~\cite{Heinrich_2025_frad, Heinrich_2025_PhD}). The reason for using this representative case is the availability of a high quality equilibrium reconstruction produced by the Integrated Data Analysis (IDA) framework \cite{Fischer_2010, Fischer_2016, Fischer_2020}. This is an H-mode D discharge with a major radius \mbox{$R_0 = 1.74\,\rm m$} at the magnetic axis, minor radius (measured along the midplane from the magnetic axis to the last closed flux surface) \mbox{$a = 0.39\,\rm m$} and an on-axis toroidal magnetic field \mbox{$B_0 = 1.81\,\rm T$}. The initial core densities are \mbox{$n_{e0} = n_{D0} \approx 6\cdot 10^{19}\,\rm m^{-3}$}, the core temperatures are \mbox{$T_{e0}\approx T_{D0}\approx 4 \,\rm keV$} and the plasma current is \mbox{$I_0 = 815\,\rm kA$}. The full profiles and flux surface geometry are displayed in figure 1 in \cite{Halldestam2025}.

The IDA equilibrium reconstruction finds a significant contribution to the plasma pressure from fast ions introduced by the Neutral Beam Injection~(NBI) heating. Since the DREAM code currently does not include a model for non-thermal ions, the pressure contribution from these ions is accounted for by combining all ions into a single ion species, with a temperature scaled to match the total ion pressure found by the equilibrium reconstruction. This implicitly assumes that the fast ions rapidly thermalise when the SPI starts and when the NBI heating is turned off. 

The boundary condition for the electric field is set at the effective wall radius $b = 0.55\,\rm m$. The resistive time scale of the wall in AUG is expected to be of the order of a few milliseconds. However, using such a value for the resistive time scale was found to lead to a significant slowing down of the current decay towards the end of the CQ, which is not observed experimentally. This may be explained by the fact that the plasma undergoes a significant vertical motion during the CQ~\cite{Heinrich_2025_PhD, Schwarz_2023}, turning a significant part of the plasma current into a halo current. As the wall resistivity may be relatively high compared to the plasma, especially for discharges with relatively small amounts of injected impurities, the wall resistivity may have a significant effect on the current decay during the later parts of the CQ. As DREAM uses a fixed flux surface geometry, this effect is not accounted for in the present model. Instead, \cite{Halldestam2025} was able to obtain a reasonable match of the experimental current evolution assuming a perfectly conducting wall, and we therefore make the same assumption here. We also note that, since the AUG simulations presented in this paper focus on the density evolution instead of the runaway dynamics, matching the current evolution very accurately is of secondary importance, as it only affects the density evolution during the recombination phase, via the Ohmic heating. 

The simulation settings related to the SPI are taken from the discharges \#40743 and \#40732. The former is a staggered injection, starting with a pure D pellet with diameter $d = 8 \,\rm mm$ and length $L=4.67\,\rm mm$ (compare~\cite{Heinrich_2024_SPI_Lab, Heinrich_2025_PhD}), followed by a Ne doped pellet. In this paper we will, however, only study the first injection stage. This discharge was selected because it had the most successful post-injection Thomson scattering measurement of the density profile following a pure D injection available from the campaign. The injection used guide tube 1~\cite{Dibon2023, Heinrich_2025_frad, Heinrich_2025_PhD} with a shattering angle $\theta _\mathrm{s}=25^\circ$. After the pellet is shattered with an impact speed $v_\mathrm{p} = 270\,\rm m/s$, the speed distribution of the shards is assumed to be Gaussian, with a mean shard speed $\langle v_\mathrm{s}\rangle$ given by the average of the initial speed and its component parallel to the bent part of the shatter tube, $\langle v_\mathrm{s} \rangle = v_0(1+\cos{\theta _\mathrm{s}})/2$, and a standard deviation of $0.2\langle v_\mathrm{s}\rangle$. This is supported by observations by \cite{Peherstorfer2022}. The coordinates of the shattering point are $(X,Y,Z) = (2.308\,\mathrm{m},-0.500\,\mathrm{m},0.328\,\mathrm{m})$ in machine coordinates with the origin at the center of the torus and $Z$ pointing vertically. The shards travel in the direction $\hat{v} = (-0.830, 0.258, -0.494)$, essentially pointing towards the magnetic axis, with a divergence angle of $\alpha = 10^\circ$. 

Discharge \#40732 was instead a single-stage injection of a pellet doped with 1.25\% Ne, with diameter $d = 7.9\,\rm mm$ and length $L=9.7\,\rm mm$. This case was chosen as an example of where the plasmoid drift is partly suppressed by a rather small Ne fraction in the pellet, which also had a reasonably successful line-integrated density measurement, despite the plasma disrupting during the injection of interest. The injection used guide tube 3 with a shattering angle $\theta_\mathrm{s} = 12.5^\circ$, injection speed $v_\mathrm{p}=230\,\rm m/s$ and coordinates of the shattering point $(X,Y,Z) = (2.307\,\mathrm{m},-0.480\,\mathrm{m},0.374\,\mathrm{m})$. The shards travel in the direction $\hat{v} = (-0.910, 0.274, -0.311)$, pointing somewhat above the magnetic axis, with a divergence angle $\alpha = 10^\circ$. The speed distribution of the shards is assumed to be Gaussian, with a standard deviation of the shard speed of $\pm 0.2\langle v_\mathrm{s}\rangle$, as for shot \#40743 described above.

For the AUG simulations, we calculate the number of shards using the model by \cite{Gebhart2020}. As described in section~\ref{sec:SPI}, the number of fragments can then differ between different samples drawn from the statistical model. In this paper, we have chosen representative samples with 177 shards for discharge \#40743 and 6 shards for discharge \#40732. Our simulations indicate that samples with different numbers of shards may result in a relative modification of the assimilation rate by typically a few tens of percent for a given value of $\Delta y$. This change is, however, not sufficient to alter the conclusions of the paper, and a similarly good agreement with the AUG experiments shown in section~\ref{sec:AUG} can still be obtained by varying $\Delta y$ within a reasonable error margin.

In discharge \#40732, the plasma disrupts a couple of milliseconds after the pellet arrives in the plasma. We emulate this by imposing enhanced transport coefficients at the onset of the disruption, with a heat diffusion coefficient of $D_W = 100\,\rm m^2/s$, a hyperresistivity parameter $\Lambda = 10^{-5}\,\rm Wb^2m/s$, an ion diffusion coefficient of $D_\mathrm{ion} = 100\,\rm m^2/s$ and ion advection coefficient of $A_\mathrm{ion} = -100\,\rm m^2/s$. These values are similar to those found to match the evolution of massive gas injection experiments in AUG in \cite{Linder2020Self}, and the value of $\Lambda$ is chosen to result in a reasonable size of the current spike in the beginning of the disruption. The onset time of the transport coefficients is set such that the time between the pellet arrival and the onset of the disruption matches the experiment, with the onset of the disruption in the experiment characterised by the minimum in the plasma current just before the current spike. For shot \#40732, this minimum occurs $1.9\,\rm ms$ after the pellet arrival. All transport coefficients then decay exponentially with a time scale $t_\mathrm{TQ,exp} = 1\,\rm ms$.

In discharge \#40743, the plasma does not disrupt during the first injection stage studied here. We therefore impose constant transport coefficients of $D_W = 4.5\,\rm m^2/s$ and $D_\mathrm{ion} = 2\,\rm m^2/s$ in order to obtain a flattening of the density and temperature profiles over a reasonable time scale. These coefficients are similar to those used in \cite{PatelMSc, Patel_2024} to simulate the same shot with the INDEX code. In cases where the plasma disrupts, the flattening of the temperature and density profiles is dominated by the major transport event, and we therefore set the transport coefficients at other times to zero to reduce the complexity of the model.

The ionization, recombination and line radiation rates are taken from the ADAS database, assuming a completely transparent plasma. We note that the densities reached in the AUG discharges studied here do not reach the $\sim 10^{22}\,\rm m^{-3}$ where the opacity to Lyman radiation is expected to play a significant role \cite{Vallhagen2022Effect}. Additional tests also show that assuming that the plasma is opaque to Lyman radiation notably delays the recombination phase compared to the experiments.

\subsection{ITER scenarios}
\label{sec:ITERSettings}
The ITER simulations presented in this paper are based on the case labeled St4$^*$ in appendix B of \cite{Vallhagen_2024}. This is a H-mode case with a core temperature of $\sim 20\,\rm keV$ and density $\sim 8\cdot 10^{19}\,\rm m^{-3}$, equally divided between deuterium and tritium. The minor radius is $a = 1.81\,\rm m$, major radius $R_0 = 6.30\,\rm m$, on-axis magnetic field $B = 5.38\,\rm T$ and the plasma current is $15\,\rm MA$. The resistive time scale of the vessel wall is set to $t_\mathrm{wall} = 500\,\rm ms$, and the boundary condition for the electric field is applied at the minor radius coordinate $b = 2.833\,\rm m$. This value of $b$ was chosen to match the magnetic field energy inside the first toroidally closed conducting wall found in JOREK simulations. The full profiles and flux surface geometry, produced by the CORSICA code \cite{Kim_2018}, are displayed in figure 1 in \cite{Vallhagen_2024}.

As a baseline case, we consider an injection starting with a pure D\footnote{The ITER disruption mitigation system plans to use H pellets, but we consider D injections here since the ablation rate presented in \cite{Parks_TSDW} contains a scaling factor fitted to D/Ne mixtures specifically.} pellet containing $1.85\cdot 10^{24}$ atoms, which is shattered into 487 shards, followed after $5\,\rm ms$ by a similar pellet but doped with $2.5\cdot 10^{22}$ Ne atoms (1.35\%). These settings were found to give the best RE mitigation performance among the $15\,\rm MA$ DT H-mode cases studied in \cite{Vallhagen_2024}. The shards are initiated at the shattering point with coordinates $(R, Z) = (8.568\,\rm m, 0.6855\,\rm m)$ and travel along the midplane with an average speed of $v_\mathrm{p}=500\,\rm m/s$, a uniform speed distribution between $v_\mathrm{p}\pm \Delta v_\mathrm{p}$ with $\Delta v_\mathrm{p}/v_\mathrm{p}=0.4$, and a divergence angle of $\alpha = \pm 10^\circ$. When plasmoid drifts are included, the width of the pellet cloud is nominally set to $\Delta y = 12.5\,\rm mm$, which is representative of the values found in simulations by \cite{Akinobu_2022} in an ITER-like setting. In addition, we consider variations of this baseline case where the pellet cloud width is increased by 50\% to $\Delta y = 18.75\,\rm mm$, and where the first pellet is doped with $5\cdot 10^{21}$ Ne atoms (0.27\%).

\begin{figure*}
    \centering
    \hspace{-1cm}
    \includegraphics[width=0.45\textwidth]{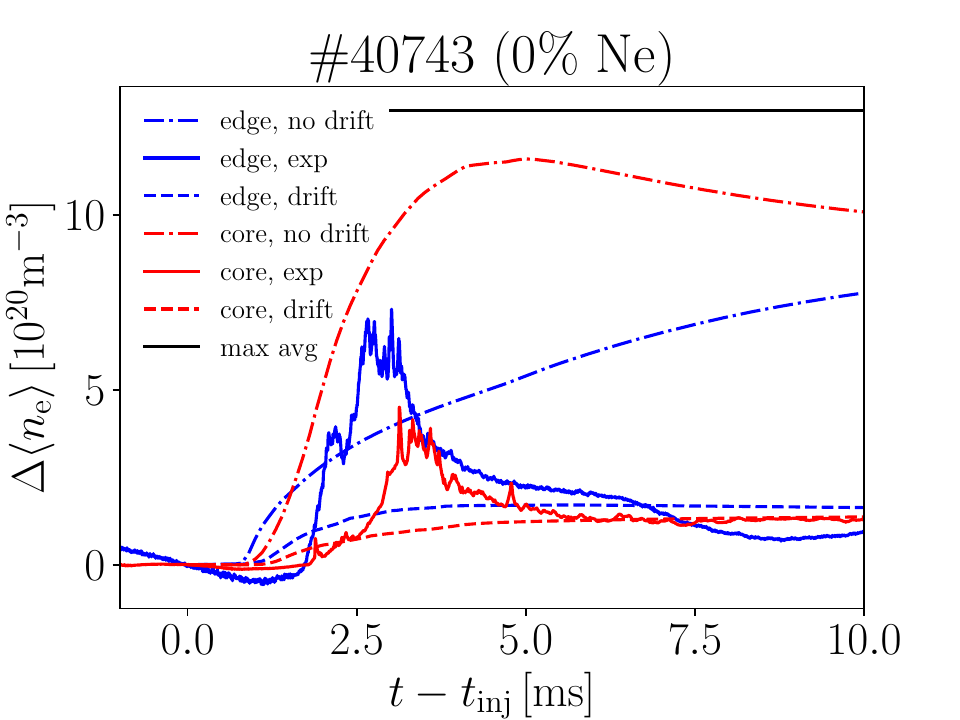}
    \put(-42,125){\large a)}
    \hspace{-0.6cm}
    \includegraphics[width=0.45\textwidth]{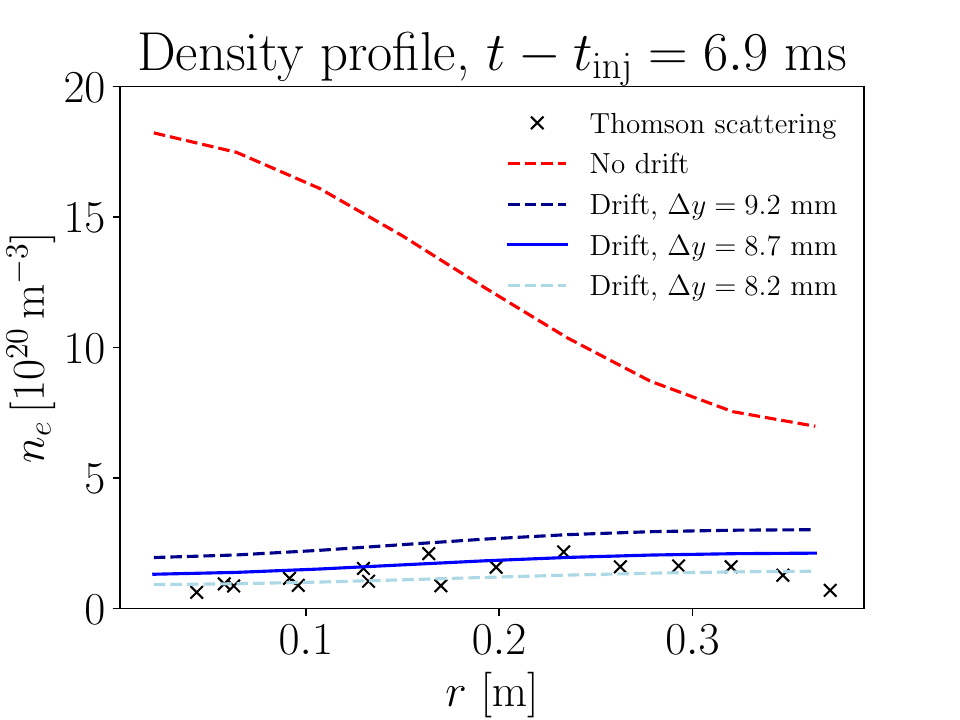}
    \put(-190,120){\large b)}
    
    \includegraphics[width=0.9\textwidth]{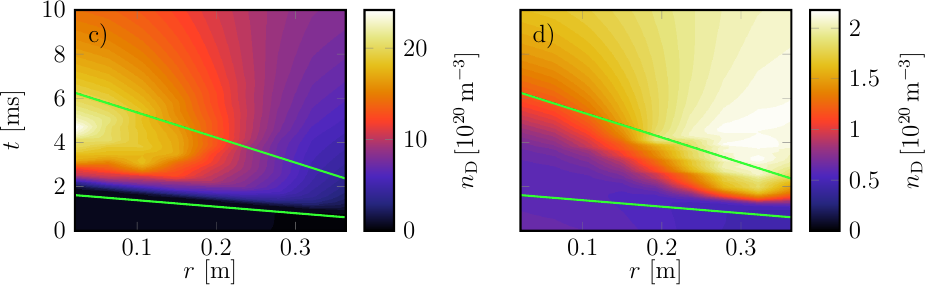}
    \caption{Simulated and measured density evolution of AUG shot \#40743. a) Line averaged density increase compared to that before the pellet arrival at the core (red) and at the edge (blue), measured in the experiment (solid) and simulated without (dash-dotted) and with (dashed) plasmoid drifts, assuming $\Delta y = 8.7\,\rm mm$. A black solid line shows the line averaged density obtained if all material in the pellet was homogeneously distributed over the plasma. b) Density profile $6.9\,\rm ms$ after the pellet arrival, measured in the experiment (`$\times$'), simulated without drift (red dashed), and simulated with drift assuming $\Delta y = 8.7\,\rm mm$ (blue solid), $\Delta y = 9.2\,\rm mm$ (black dashed) and $\Delta y = 8.2\,\rm mm$ (light blue). c) Simulated radial evolution without the drift and d) with drift assuming $\Delta y = 8.7\,\rm mm$. In c-d) solid green lines indicate the trajectory of the fastest and slowest shards.}
    \label{fig:40743}
\end{figure*}

In the baseline case, the plasma is assumed to disrupt when the temperature drops below $10\,\rm eV$ anywhere inside the flux surface on which the safety factor $q=2$ (referred to as a ``late'' TQ onset). We then impose a hyperresistivity of $\Lambda = 0.628\,\rm Wb^2m/s$, giving a current spike of $\sim 0.5-1\,\rm MA$ ($\sim 3-7\%$ of the plasma current) depending on the scenario, which is consistent with e.g. MHD simulations and experiments of massive gas injections at JET \cite{Gerasimov_2020, Nardon_2021} and SPI experiments at AUG \cite{Halldestam2025}. We also impose a Rechester-Rosenbluth-type heat transport with magnetic perturbation amplitude $\delta B/B = 1.85\cdot 10^{-3}$, and an RE transport based on the model in \cite{Svensson2021Effects} (see section~\ref{sec:current}) with the same value of $\delta B/B$. These transport coefficients are nominally active for \mbox{$t_\mathrm{TQ} = 3\,\rm ms$}, and the value of $\delta B/B$ is adapted such that the temperature decays to around $200\,\rm eV$ at the end of the transport event in a separate simulation with transport as the only loss mechanism. At this temperature, radiative losses tend to become dominant compared to transported losses. We also consider a variation of this scenario where the transport event starts when a shard travelling at the average shard speed would reach the $q = 2$ flux surface (referred to as an ``early'' TQ onset) and the aforementioned transport coefficients are active for $t_\mathrm{TQ} = 1\,\rm ms$. In that case, the magnetic perturbation amplitude is rescaled to $\delta B/B = 3.20\cdot 10^{-3}$ in order to reach the same final temperature with transport as the only loss mechanism.

At the start of the transport event, we also impose ion transport coefficients peaking at $A_\mathrm{ion} = - 2000\,\rm m/s$ and $D_\mathrm{ion} = 4000\,\rm m^2/s$ and decaying exponentially with a time constant $t_\mathrm{ion} = 0.5\,\rm ms$. These values were chosen in order to achieve an ion mixing on a similar time scale as observed in 3D JOREK simulations of ITER disruptions \cite{Hu2021Radiation}. 
Finally, the ionization, recombination and line radiation coefficients are taken from the AMJUEL database for hydrogenic species to account for opacity to Lyman radiation (see section \ref{sec:SPI}).

\section{Simulation results}
\label{sec:results}

\subsection{AUG simulations and experimental comparison}
\label{sec:AUG}

The plasmoid drift is expected to be most pronounced in pure D injections, due to the lack of radiative losses decreasing the plasmoid pressure. In all AUG cases studied here, the simulated RE current was negligible ($<10^{-5}\,\rm A$), consistent with the experiments, and we therefore focus our analysis on the injection dynamics. The simulated and experimental density evolution for the first, pure D, injection stage of discharge \#40743 is shown in figure~\ref{fig:40743}. The line averaged density increase compared to that before the pellet arrival in the core ($R = 1.79\,\rm m$) and the edge ($R=1.20\,\rm m$) is shown in figure~\ref{fig:40743}a. Here, we see that the measured line averaged density is significantly smaller than would be obtained if all the pellet material was evenly distributed over the plasma, with the long-term core line averaged density being $90\%$ lower. A similarly large discrepancy is found compared to a simulation neglecting the plasmoid drift, indicating that the plasmoid drift plays a major role in this case. 

The measured line averaged density is calculated by dividing the measured line integrated density from $\rm CO_2$ laser interferometry~\cite{Mlynek2012,Mlynek2017} with the chord length calculated by the equilibrium reconstruction immediately before the pellet arrived at the plasma. This is done to avoid issues as the equilibrium reconstruction becomes inaccurate during the off-normal rapid density rise. The simulated line averaged density is normalised to the cord length of the corresponding line for the flux surface geometry used in the simulations (see section \ref{sec:AUGsettings}).

The simulation agrees reasonably well with the experiment when including the plasmoid drift with $\Delta y = 8.7\,\rm mm$. This value of $\Delta y$ was chosen such that the simulated core line averaged density best matches the corresponding long-term measured line averaged density. It also reproduces the generally higher long-term line averaged density at the edge reasonably well. Shortly after the injection, the measured line averaged density shows a somewhat noisy temporal peak, which is not seen in the simulation. This may be explained by plasmoids temporarily passing the line of sight before homogenising. Note that the probability of such structures passing the line of sight is significantly increased by the poloidal shear of such structures passing through the scrape-off layer, due to the large $E\times B$-rotation of the plasma there. Reproducing such features would require the temporal evolution of the plasmoid drifts as well as the shear in the scrape-off layer to be resolved in the simulation. It is therefore not expected that the temporal peak in the line averaged density is visible in the simulation even if the long-term evolution agrees with the experiment.

The long-term agreement with the experiment is confirmed by the comparison of the simulated density profile to the Thomson scattering measurement~\cite{Kurzan2011, Murmann1992}, shown in figure~\ref{fig:40743}b. The measured and simulated density profiles have a similar size and shape, with a $\lesssim 2$ times as high density at the edge compared to the core.
Note, however, that the simulated density profile is very sensitive to the value of $\Delta y$, with a difference in $\Delta y$ of only $\pm 0.5\,\rm mm$ altering the density profile by $\sim$30-50\%. 

The full simulated radial density evolution is shown in figure~\ref{fig:40743}c without the drift and in figure~\ref{fig:40743}d including the drift with $\Delta y = 8.7\,\rm mm$. Without the drift, the start of the density build-up clearly coincides with the arrival of the pellet shard plume. The build-up is also larger in the core, due to the higher temperature and smaller flux surface cross section area there, and the core density starts to decrease due to the transport after the passing of the shard plume. With the drift included, the density build-up lags behind the front of the shard plume, and the initial deposition profile is quite strongly peaked at the edge before it is flattened by the transport. Notably, barely any material is deposited directly in the core, but the build-up of the core density is quite strongly reliant on the diffusive transport.

Adding a small Ne doping to the pellet strongly enhances the radiation losses from the plasmoids, reducing the plasmoid pressure and hence the driving mechanism of the drift. We emulate this effect by reducing the plasmoid temperature from 30 eV to 5 eV as described in section~\ref{sec:SPI}. This leads to a strong reduction in the simulated drift losses, as shown shortly.

\begin{figure}
    \centering
    \includegraphics[width=\linewidth]{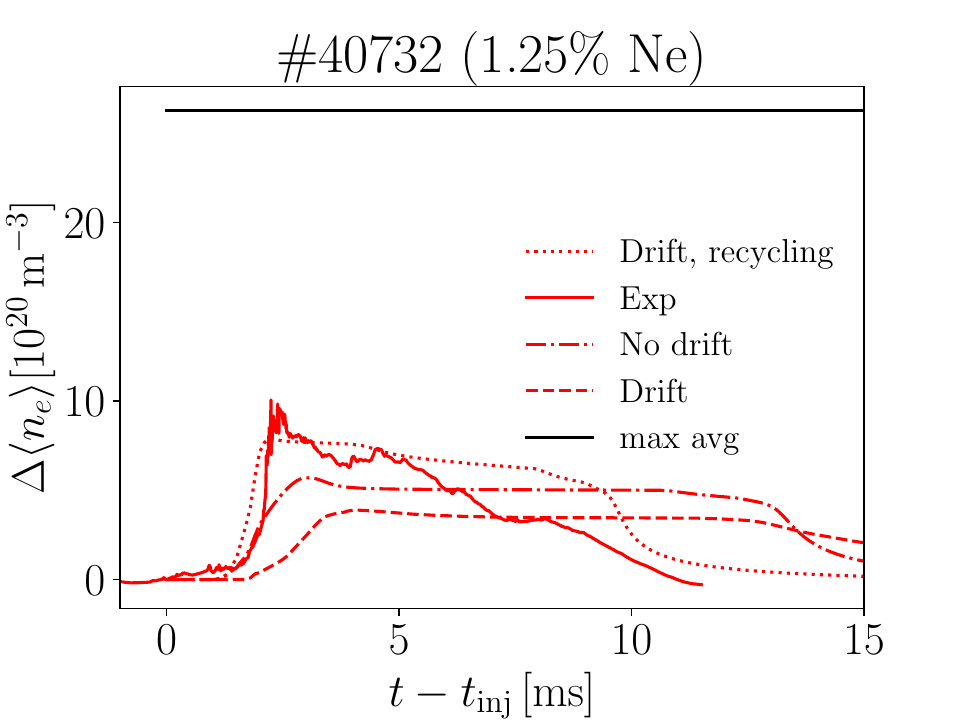}
    \caption{Central line averaged density evolution for AUG discharge \#40732, comparing the experimental result (solid) to simulations without drifts (dash-dotted), with drifts passing freely through the plasma edge (dashed) and with drift assumed to stop at the plasma edge (dotted). A black solid line shows the line averaged density obtained if all material in the pellet was homogeneously distributed over the plasma.}
    \label{fig:40732}
\end{figure}

Measured and simulated core line averaged densities for discharge \#40732, with an injection of a pellet with 1.25\% Ne (more details are given in section~\ref{sec:AUGsettings}), are shown in figure~\ref{fig:40732}. In the simulations including drifts, we set $\Delta y = 10\,\rm mm$, but note that the results are not quite as sensitive to $\Delta y$ as before; using $\Delta y=8.7\,\rm mm$ as in figure \ref{fig:40743} decreases the long-term assimilation rate of the case labeled ``Drift'' in figure \ref{fig:40732} (the different labels will be described shortly) by 26\%, and does not cause any other essential changes. 

\begin{figure*}
    \centering
    \includegraphics[width=\textwidth]{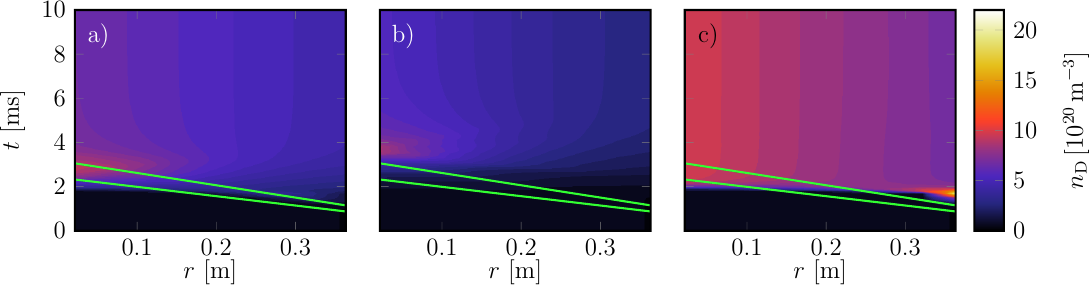}
    \caption{Simulated radial density evolution in AUG discharge \#40732, a) without drifts, b) with drifts passing freely through the plasma edge and c) with drifts assumed to stop at the plasma edge. The solid green lines indicate the trajectory of the fastest and slowest shards.}
    \label{fig:40732radial}
\end{figure*}

Here, we only study the core line averaged density, as the motion and reshaping of the plasma following the disruption can have a major impact on the chord length passing through the plasma at the edge. This would necessitate using a continuously updated chord length after the disruption, which the equilibrium reconstruction could not reliably provide. The simulation also uses a fixed flux surface geometry, and therefore changes in the line averaged density caused by changes in the flux surface geometry can not be modelled. This issue is also present in the core, but for geometrical reasons it is not as pronounced there. The interferometer measurement in the core was disturbed by fringe jumps, as evidenced by the final density differing from zero by an amount corresponding to 2 fringes. The likely time of the fringe jumps could, however, be narrowed down to a short period shortly after the onset of the disruption. In figure \ref{fig:40732}, we have therefore corrected for the fringe jump by shifting the interferometer data by 2 fringes $2.25\,\rm ms$ after the pellet arrival.

The measured density rise is still $\sim 3$ times smaller than if all the injected material was homogeneously spread across the plasma, but the ratio is still $\sim3$ times larger than the long term density rise for the pure D injection in discharge \#40743 (disregarding the initial peak present in this case). The difference is even more striking when comparing to the simulation without the drift; even in this case the measured density rise is higher than the simulated one for a significant amount of time. Part of this could be caused by a similar peaking due to temporarily over-pressurized structures passing the line of sight, but these are expected to be rapidly homogenised when the plasma disrupts at $t-t_\mathrm{inj} = 1.9\,\rm ms$. It is also not possible to distinguish a temporal peak, which is clearly separated from the long-term behaviour as in discharge \#40743. In addition, the density drops earlier due to recombination in the experiment than in the simulation without drift, indicating a higher assimilation in the experiment which leads to a faster radiation cooling. Altogether, these observations indicate a major reduction of the drift compared to the pure D injection in discharge \#40743.

The reduced impact of the drift is found also when comparing the simulation with and without the drift. As the drift leads to a reduction of the assimilation rate, the simulation with the drift now agrees less well with the experiment than the simulation without the drift, but the difference is modest; the difference in the maximum density rise is only 32\% between the simulations with and without the drift. The simulated radial density evolution, shown in figure~\ref{fig:40732radial}, is also qualitatively similar with and without the drift. In both figure~\ref{fig:40732radial}a (without drift) and b) (with drift), the direct density rise follows fairly shortly after the passing of the front of the shard plume and is highest in the core, before the profile is flattened by the transport event.

The agreement with the experiment can, however, be improved by modifying the assumption about the plasmoid dynamics around the scrape-off layer. The strong $E\times B$ rotation of the plasma in the scrape-off layer creates a shear which effectively tears apart over-pressurised structures. This phenomenon has previously been studied in the case of blob transport \cite{Zholobenko_2021}, and could potentially stop or delay the loss of the drifting material until it can be recycled during the homogenisation caused by the transport event. When the magnetic field lines become stochastic and the plasma pressure rapidly drops, it may also be possible for material in the vacuum region which has not yet been pumped out to get recycled (this situation is similar to the penetration of injected gas, modelled in e.g. \cite{Nardon_2017}).

We have performed simulations emulating such an effect (labeled ``Drift, recycling'' in figure \ref{fig:40732}) by assuming that the material whose drift reaches the plasma edge is stopped in the outermost grid cell in the simulation ($4.3\,\rm cm$ wide). This material is then redistributed over the plasma during the transport event, as shown in figure~\ref{fig:40732radial}c. The corresponding maximum core line averaged density increase now matches the experiment well (see figure \ref{fig:40732}, dotted line), and the time scale of the decrease due to transport and recombination is also reasonably well reproduced. The fact that the line averaged density is now higher than in the simulation without drifts can be explained as the drift now shifts the cooling of the background plasma to larger radii, that is behind the front of the shard plume. With the present boundary condition for the drift, such a shift increases the ablation without leading to any material being lost from the plasma\footnote{Note that this means that applying the boundary condition used here to the pure D case shown in figure \ref{fig:40743} would make the agreement with the experiment even worse than for the simulation without drifts included. This boundary condition was therefore not considered there.}. This could be interpreted as an indication that plasmoids reaching the edge do not necessarily escape the plasma immediately, which could potentially have a significant effect on the assimilation rate. However, it remains to be confirmed that an accumulation at the edge actually takes place, by e.g. performing experiments with more reliable edge density measurements.

\subsection{ITER simulations}
\label{sec:ITER}

\begin{table*}[]
    \centering
    \begin{tabular}{c|c|c|c|c|c|c|c|c|c}
      \# & Drift & $\Delta y\,\rm [mm]$ & Ne 1 & TQ onset & $t_\mathrm{TQ}\,\rm [ms]$ & Assim. 1 [\%]& Assim. 2 [\%] & $I_\mathrm{RE}^\mathrm{seed}\,\rm [mA]$& $I_\mathrm{RE}^\mathrm{repr}\,\rm [MA]$\\\hline
        1 & No  & 12.5 & 0 & Late & 3 & 94 & 41 & 14.0 & 5.98\\
        2 & Yes & 12.5 & 0 & Late & 3 & 4.1 & 79 & 42.8 & 6.14\\
        3 & Yes & 18.75 & 0 & Late & 3 & 59 & 44 & 13.6 & 5.77\\
        4 & Yes & 12.5 & $5\cdot 10^{21}$ & Late & 3 & 71 & 23 & 14.0 & 5.69\\
        5 & Yes & 12.5 & 0 & Early & 1 & 4.1 & 34 & 400 & 7.15\\
        6 & Yes & 12.5 & $5\cdot 10^{21}$ & Early & 1 & 71 & 9.7 & 4.4 & 5.99\\
    \end{tabular}
    \caption{Settings and figures of merit for the ITER simulations presented in this paper. ``Ne 1'' refers to the number of Ne atoms in the first injection stage, and ``Assim. 1'' and ``Assim. 2'' refer to the assimilation rate of the first and second injection stage, respectively. $I_\mathrm{RE}^\mathrm{seed}$ is defined as the sum of the RE current remaining after the transport event and the seed generated after the transport event. $I_\mathrm{RE}^\mathrm{repr}$ is defined as the RE current when 95\% of the total plasma current is carried by REs. The other parameters are defined in sections \ref{sec:model} and \ref{sec:ITERSettings}.}
    \label{tab:settings_results}
\end{table*}

\begin{figure*}
    \centering
    \includegraphics[width=0.9\textwidth]{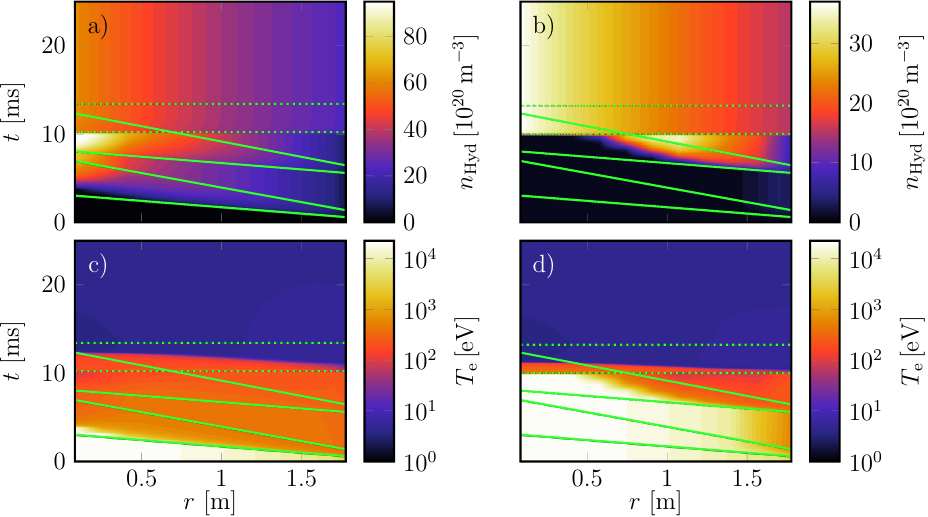}
    \includegraphics[width=0.45\textwidth]{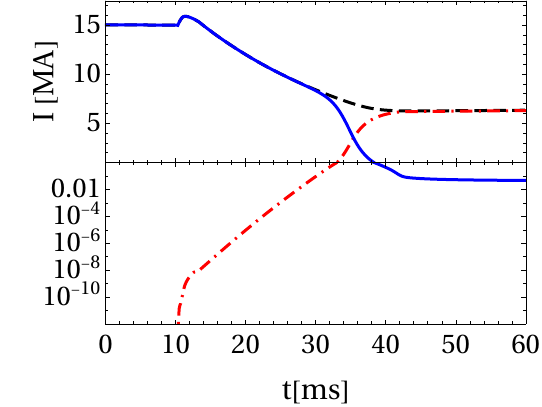}
    \put(-25,145){\large e)}
    \includegraphics[width=0.45\textwidth]{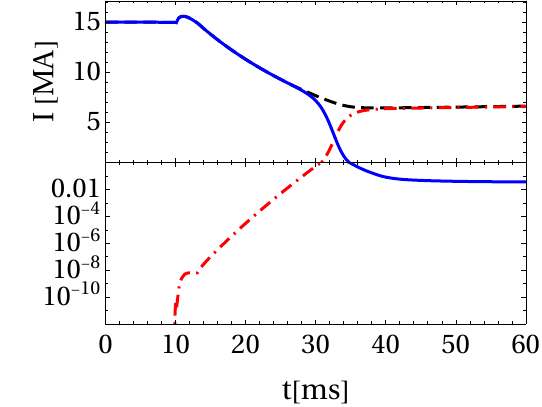}
    \put(-25,145){\large f)}
    \caption{Evolution of plasma parameters for the ITER case with baseline parameters. The left column shows the a) density, c) temperature e) and current evolution without plasmoid drifts, and the right column (b,d and f) shows the corresponding results with plasmoid drifts assuming a cloud thickness of $\Delta y = 12.5\,\rm mm$. In a-d) solid green lines indicate the trajectory of the fastest and slowest shards for each injection and the temporal boundaries of the transport event are shown with dotted green lines.}
    \label{fig:St4}
\end{figure*}

\begin{figure*}
    \centering
    \includegraphics[width=0.9\textwidth]{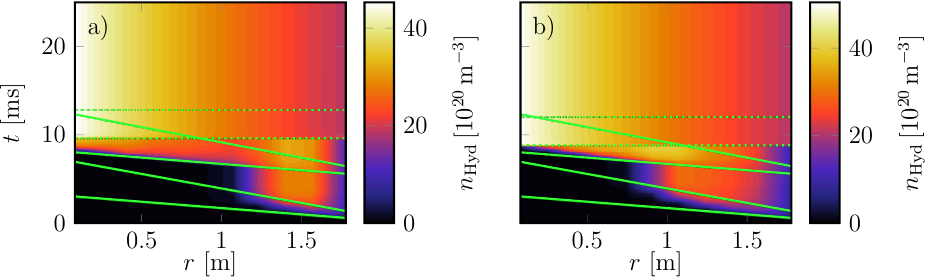}
        \caption{Density evolution for the ITER case with plasmoid drifts assuming a) a plasmoid thickness $\Delta y = 18.75\,\rm mm$ and a pure D pellet in the first injection stage, and b) a plasmoid thickness $\Delta y = 12.5\,\rm mm$ and a doping with $5\cdot 10^{21}$ Ne atoms in the first injection stage. Solid green lines indicate the trajectory of the fastest and slowest shards for each injection and the temporal boundaries of the transport event are shown with dotted green lines.}
    \label{fig:St4_delta_y_18_75}
\end{figure*}

\begin{figure*}
    \centering
    \includegraphics[width=0.9\linewidth]{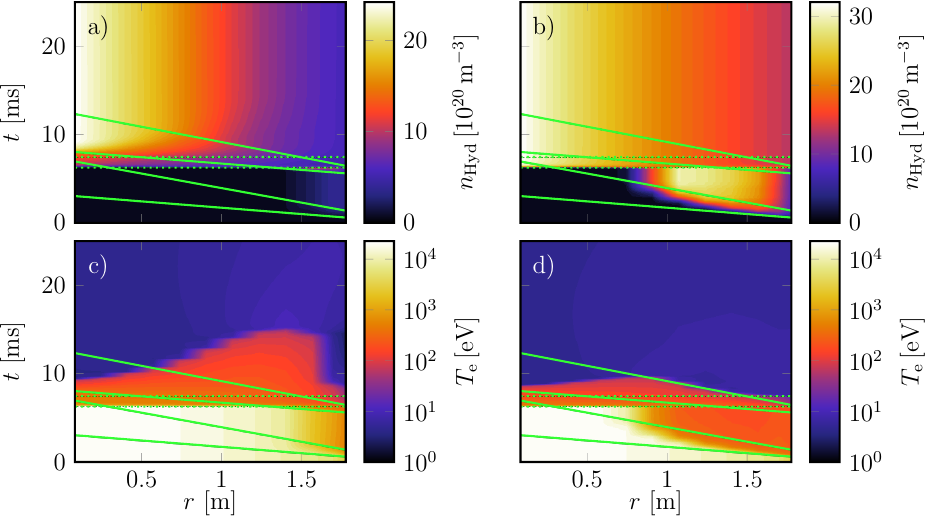}
    \includegraphics[width=0.45\textwidth]{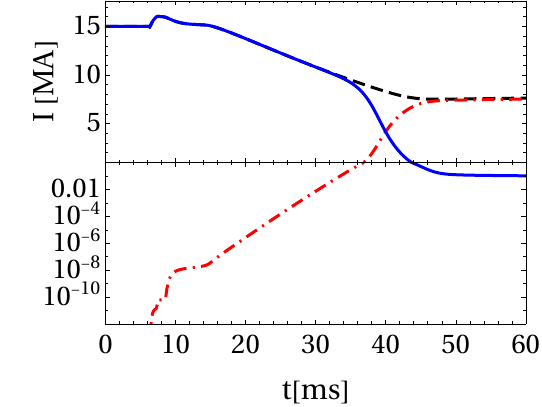}
    \put(-25,145){\large e)}
    \includegraphics[width=0.45\textwidth]{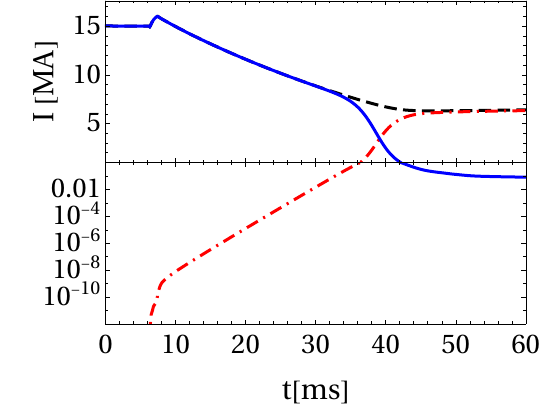}
    \put(-25,145){\large f)}
    \caption{Evolution of plasma parameters for the ITER case ran with a $1\,\rm ms$ TQ with an early onset, both including plasmoid drifts with a plasmoid thickness $\Delta y = 12.5\,\rm mm$. The left column shows the a) density, c) temperature e) and current evolution using a pure D pellet in the first injection stage, and the right column (b,d and f) shows the corresponding results with an Ne content of $5\cdot 10^{21}$ atoms in the first pellet. In the latter case we assume a reduced plasmoid temperature of $T=5\,\rm eV$ while still assuming that the TQ onset takes place when a shard travelling at the average speed in the second injection passes the $q=2$ flux surface. In a-d) solid green lines indicate the trajectory of the fastest and slowest shards for each injection and the temporal boundaries of the transport event are shown with dotted green lines.}
    \label{fig:St4_bad_TQ}
\end{figure*}

The typical distance drifted is expected to be longer in an ITER-like setting, as larger pellets and hotter plasmas lead to a more intense ablation and higher plasmoid pressures. It is therefore expected that plasmoid drifts may have an even larger impact on the assimilation rate of pure hydrogenic SPIs in ITER compared to AUG. This is investigated in figures~\ref{fig:St4}-\ref{fig:St4_bad_TQ}, comparing the disruption dynamics with different drift settings for variations of the baseline ITER case detailed in section~\ref{sec:ITERSettings}. The main settings and resulting figures of merit for the cases studied in this section are also summarised in table \ref{tab:settings_results} (the settings not specified there are always the same as in the baseline case).

We assume here all material can freely drift across the plasma edge, as no conclusive evidence for a significant reduction of the drift at the edge could be drawn from the AUG simulations. Moreover, for Ne-doped pellets, where the AUG simulations indicated a significant reduction of the drift across the boundary, the calculated drift distances in the ITER cases shown in figure \ref{fig:St4}-\ref{fig:St4_bad_TQ} are typically not larger than a few decimeters. Thus, since this is $\sim10$ times smaller than the minor radius in ITER, the fraction of material ablated from such pellets that reaches the edge is relatively small, as will be shown shortly.

Without drifts (case 1 in table \ref{tab:settings_results}), the direct density increase is peaked in the core for both the first and second injection stage, as shown in figure~\ref{fig:St4}a. As a result, the vast majority of the pellet, 94\%, is assimilated in the first injection stage. The assimilated material cools the plasma to temperatures in the $10^2-10^3\,\rm eV$-range, as shown in figure~\ref{fig:St4}c, reducing the assimilation rate of the second injection stage to 41\%. The radiation losses from the Ne-doping also introduce another cooling mechanism in addition to the dilution, which further reduces the ablation rate. The cooling rate remains relatively low for $t\approx 3-12 \,\rm ms$, despite the Ne radiation losses, until the shard plume reaches the plasma core, at which point the transport event is triggered and the temperature rapidly drops to the $\sim 5\,\rm eV$ range. 

The phase with a relatively slow temperature decay gives the electron momentum distribution time to thermalise at an intermediate level, effectively suppressing the hot-tail RE seed mechanism, as previously found in \cite{Vallhagen2022Effect}. The high density increase also effectively suppresses the Dreicer and tritium seed mechanisms. The high energy of the gamma photons from the activated wall, however, makes it infeasible to suppress the Compton seed, enabling the generation of a representative seed (defined as the sum of the RE current remaining after the transport event and the seed generated after the transport event) of $14.0\,\rm mA$, mostly generated by the Compton mechanism. The strong avalanche gain resulting from the high plasma current in ITER then amplifies this seed to a macroscopic representative RE current (defined as the RE current when 95\% of the total plasma current is carried by REs) of $5.98\,\rm MA$, see figure~\ref{fig:St4}e.

With drifts included (case 2 in table \ref{tab:settings_results}), most of the material from the first pellet is lost from the plasma, as shown in figure~\ref{fig:St4}b; the assimilation rate is now only 4.1\%. The material which is deposited is very localised to the edge, leading to a moderate but visible temperature drop there (see figure~\ref{fig:St4}d) but the temperature is left essentially unaffected in the rest of the plasma. This enables a faster ablation of the second pellet than in the case without drifts included. The ablation is further enhanced by the shift of the plasma cooling behind the front of the shard plume by the drift. However, due to the Ne doping, the distance drifted is not long enough to expel a large amount of material from the plasma, resulting in a high assimilation rate of 79\% for this pellet. The total hydrogen assimilation rate is therefore only about 40\% lower with drifts compared to without.

The low assimilation rate of the first pellet leads to a more abrupt cooling of the plasma once the transport event is triggered, especially in the plasma core. There is, however, an intermediate slow cooling phase in most of the plasma, and by the time the temperature reaches low values in the core there is already a significant density rise there due to transport, which effectively damps the hot-tail seed generation. The lower density, however, enables the tritium decay to contribute to the RE seed resulting in a somewhat larger characteristic seed of $42.8\,\rm mA$. The lower frictional drag resulting from the lower hydrogenic density also enables a somewhat faster avalanche. The overall difference in the RE generation compared to the case without drifts is, however, moderate; the representative RE current in this case is $6.14\,\rm MA$.

Although the above results indicate that the drifts might make the first injection stage ineffective, they may potentially underestimate the assimilation rate. In particular, the results are sensitive to the value of $\Delta y$, which was also shown for the AUG cases in figure~\ref{fig:40743}b. Increasing the baseline value by 50\% to $\Delta y = 18.75\,\rm mm$, as in case 3 in table \ref{tab:settings_results}, leads to the density evolution shown in figure~\ref{fig:St4_delta_y_18_75}a. The assimilation rate of the first pellet is now 59\%, and the assimilation rate for the second pellet is 44\%. The representative RE current of this case was $5.77\,\rm MA$. Notably, this is slightly lower than the previous cases in figure \ref{fig:St4}e-f, both with and without drifts. This may be explained as the amount of assimilated amount of hydrogen is closer to the optimal value, where the frictional drag is the highest without leading to a substantial recombination during the CQ which increases the avalanche growth rate, as described in \cite{Vallhagen2020Runaway}.

If $\Delta y$ is kept at the baseline value of $12.5\,\rm mm$, a similar result can also be obtained by adding a small amount of Ne doping to the first pellet. The density evolution of such a case, number 4 in table \ref{tab:settings_results}, is shown in figure~\ref{fig:St4_delta_y_18_75}b, where the first pellet is doped by $5\cdot 10^{21}$ Ne atoms (0.27\%). This amount of Ne atoms is not enough to trigger the transport event before the second pellet has reached far into the plasma, but is enough to substantially suppress the plasmoid drift. The assimilation rates are now 71\% for the first pellet and 23\% for the second. The representative RE current of this case is $5.69\,\rm MA$. Additional tests show that the results are insensitive to the amount of Ne atoms in the first pellet, assuming the plasmoid temperature remains at $5\,\rm eV$, in the range $\lesssim 10^{22}$, above which the timing of the onset of the transport event starts to become significantly affected.

The differences in the density evolution in the above cases have only resulted in a minor difference in the representative RE current. The difference may be larger (although still not dramatic) in cases with less favourable TQ conditions. Such a case, number 5 in table \ref{tab:settings_results}, is studied in figure~\ref{fig:St4_bad_TQ}, assuming an early transport event with a duration of $t_\mathrm{TQ} = 1\,\rm ms$ (see section~\ref{sec:ITERSettings}). Figure~\ref{fig:St4_bad_TQ}a shows the density evolution of such a case including drifts with $\Delta y = 12.5\,\rm mm$ and a pure D pellet in the first injection stage. Here we have also added a small heat transport of Rechester-Rosenbluth type with $\delta B/B=3\cdot10^{-4}$ after the main transport event, to prevent the formation of unrealistically thin hot filaments such as those described by \cite{Putvinski1997, Feher2011}. The first injection stage evolves as in figure~\ref{fig:St4}a. The assimilation rate of the second pellet is again much larger, 34\%, but is now limited by the early onset of the transport event, which cools the plasma and slows down the ablation. As the pellet shards now only have up to $0.7\,\rm ms$ to ablate before the onset of the transport event, a significant part of the assimilation takes place after the transport event, despite the relatively low temperature. This leads to a more peaked profile than with the late onset of the transport event, as a significant part of the assimilated material is not homogenised by the transport.

Despite the early and fast transport event, the low assimilation prevents a complete thermal collapse of the plasma in direct connection to the transport event, see figure~\ref{fig:St4_bad_TQ}c, again resulting in a negligible hot-tail seed. The low assimilation rate, however, enables the generation of a tritium RE seed, leading to a total seed of $0.4\,\rm A$. In combination with the low frictional drag, this leads to a relatively high RE current of $7.15\,\rm MA$ (see figure~\ref{fig:St4_bad_TQ}e).

With these TQ conditions, increasing the assimilation of the first pellet may have a significantly larger effect on the final RE current than with more favourable TQ conditions. Figure~\ref{fig:St4_bad_TQ}b shows the density evolution for the same case as above but with an Ne doping of $5\cdot 10^{21}$ atoms in the first pellet (case 6 in table \ref{tab:settings_results}). In this case it was not necessary to include any heat transport after the main transport event to avoid the formation of unrealistically thin hot filaments, and this heat transport was therefore omitted (we have confirmed that including it only had a very minor impact on the result in this case). As in figure~\ref{fig:St4_delta_y_18_75}b, the assimilation rate of the first pellet is now 71\%. The cooling from the first pellet (see figure~\ref{fig:St4_bad_TQ}d) reduces the assimilation rate of the second pellet to 9.7\%, but the combined final density is still 2.3 times higher at the edge and 1.3 times higher in the core. This extra density reduces the RE seed to $4.4\,\rm mA$, and also slows down the avalanche rate somewhat. The resulting representative final RE current is $5.99\,\rm MA$.
\section{Discussion}
\label{sec:discussion}
The plasmoid drift model implemented in DREAM appears to be able to explain experimentally observed assimilation rates and density profiles during SPI operations reasonably well. This has been exemplified here for two AUG discharges both with and without Ne doping. Good agreement with the experimental density measurements was found with values of the free parameters of the drift model which are consistent with previous measurements and simulations. The agreement could likely be improved by adjusting the assumed transport coefficients, although such fine-tuning is besides the point of this study. 
The predictive power of the drift model is, however, limited by its sensitivity to its free parameters. Representative values of the free parameters may be determined based on measurements and simulations available in the literature, as well as estimates based on the ionization and radiation characteristics of the plasmoids. However, the values of the free parameters, and their dependence on the background plasma and SPI parameters, as well as their interdependence, are not known well enough to narrowly constrain the model prediction. This applies in particular to the plasmoid width $\Delta y$. A power law scaling of the distance drifted $\Delta R$ with $\Delta y$ can be derived in the simplified case considered in equation (A5) in \cite{Vallhagen2023Drift} and \ref{app:Delta_y}. Here it assumed that the cloud has a long pressure equilibration time compared to the drift duration, and a rapid field line-parallel expansion. In this case it can be shown that the distance the plasmoid drifts scales as $\Delta R \sim \Delta y^{-4}$ or $\Delta R \sim \Delta y^{-4.5}$, depending on if the detachment of the cloud from the pellet is dominated by the pellet speed or the drift acceleration. 

This strong dependence is reflected in figure~\ref{fig:40743}b where a small variation in $\Delta y$ of $0.5\,\rm mm$, which is well within the error margin of $\Delta y$, was found to have a significant impact on the density evolution. Comparing figure~\ref{fig:St4}b and figure~\ref{fig:St4_delta_y_18_75}a, we see that a 50\% increase in $\Delta y$ could alter the assimilation rate of a pure D SPI by as much as an order of magnitude. Developing a self-consistent, accurate model for predicting $\Delta y$ could therefore greatly improve the predictive power of the model. In this context it is also relevant to consider the exact definition of the plasmoid width for a realistic plasmoid; the present model assumes homogeneous plasmoids with sharp edges, while a realistic plasmoid will have a smooth density variation.

The plasmoid drift model also includes several other simplifications which may affect the simulation results. This includes neglecting the striated structure of the drifting material (described in section \ref{sec:introduction}), and neglecting the interactions of plasmoids from different shards. These effects may alter the pressure variation in the stream of drifting material, thereby affecting the drift acceleration. The 3D simulations with JOREK presented in \cite{Hu2024} indicated that if the plasmoids from different shards have a substantial overlap, the ablated material from all shards may behave as a single cloud, making the drift motion insensitive to the plasmoid width of individual shards. These simulations, however, assumed an extended spread of the ablated material around the shards, for numerical reasons (a gaussian shape with a toroidal spread of $0.3\,\rm rad$ and a poloidal spread of $8\,\rm mm$). With the number of shards considered here, the shards are typically separated by a few centimeters, which is likely to reduce but not eliminate the overlapping. Moreover, the shards may interact with each other even if they do not physically overlap, by altering the electrostatic potential on the field lines connecting them. Such variations in the electrostatic potential may affect the Ohmic currents flowing from the end caps of the plasmoids, thus affecting the charge separation inside the plasmoid and the corresponding drift velocity. Accounting for these effects in detail will likely be very complicated and computationally expensive, but it might be possible to develop a sufficiently accurate model based on a statistical approach.

Another uncertainty of the present model is the behaviour of the drifting material when it passes the plasma edge. While the measurements from AUG discharge \#40743 could be reproduced assuming the material drifts freely out of the plasma, the reproduction of discharge \#40732 was notably improved when assuming that the drift stops at the plasma edge (see figure~\ref{fig:40732}). This may be explained by the shear in the scrape-off layer, corresponding to the potential gradient there, tearing apart the plasmoids and reducing the excess pressure. Another mechanism which may contribute to slowing the drift is the passing of low-order rational flux surfaces, where the connection length for the Ohmic current flowing from the end caps of the plasmoids is much lower than on irrational flux surfaces, accelerating the drain of the charge separation inside the plasmoids \cite{Pegourie2006Homogenization}. Alternatively, the drifting material may leave the plasma but be recycled before being pumped out of the vessel, perhaps aided by the drop in the plasma pressure and magnetic field stochastisation during the disruption. 

Explaining the dynamics of drifting material close to the plasma edge, and determining under which conditions a significant amount of material is stopped or recycled, may be of importance to accurately predict how much material is actually lost from the plasma due to the drift. The results presented here do, however, not contain any indications that these effects are significant for the assimilation of pure D pellets, which are most affected by the drifts. This may be related to the fact that the pure D shards drift faster, and therefore spend less time on low-order rational flux surfaces and in the scrape-off layer. A temporary slowing down and shearing of the plasmoids could however explain the early temporal peak in the line averaged density measurements observed in figure \ref{fig:40743}a.

Nevertheless, our results corroborate the expectation that pure hydrogenic SPIs are strongly affected by plasmoid drifts, while doped pellets are only rather moderately affected. This is consistent with previous findings at e.g. the DIII-D and JET tokamaks \cite{Lvovskiy2024, Kong_interpretive}. The difference is clearly seen in the studied AUG cases, and may be even more pronounced in ITER, depending on the scenario. Despite the high model sensitivity to $\Delta y$, our results indicate that a substantial part of pure hydrogenic pellet material will be lost due to the drifts in a $15\,\rm MA$ H-mode ITER scenario, perhaps even rendering the first injection stage of the staggered scheme considered here completely ineffective. It is, however, possible to assimilate a sizeable fraction of such a pellet assuming a plausible value of $\Delta y$. 

Our results also indicate that the losses from pure hydrogenic pellets injected in ITER may be significantly reduced, although perhaps not eliminated, by adding a small amount of Ne doping. For this to work, it is, however, required that the small amount of Ne doping is sufficient to keep the plasmoid temperature in the $\sim 5\,\rm eV$-range, which has to be verified experimentally or with self-consistent modelling of the plasmoid temperature. It is also required that the doping does not significantly reduce the time between the injection and the onset of the transport event. When assuming a late onset (figure~\ref{fig:St4_delta_y_18_75}b), the Ne added to the first injection stage did not significantly accelerate the onset of the transport event. When assuming an early onset (figure \ref{fig:St4_bad_TQ}), we chose to trigger the transport event only based on shards from the second injection. Given the crudeness of our model for triggering the transport event, the possibility of adding Ne doping to the first injection stage without causing a too early TQ should be verified with 3D MHD simulations and experiments.

The RE currents calculated in the ITER cases considered here have been rather insensitive to the dynamics of the first injection stage, despite strong variations with the assumptions for the drift. This is partly explained by the fact that a low assimilation rate of the first injection stage leaves the plasma hotter when the second pellet arrives, increasing the assimilation rate of the second pellet. Another major reason is the self-regulation of the RE current via the induced electric field when the RE current becomes comparable to the total plasma current; a slower RE generation allows for a higher induced electric field, which partly compensates for the slower RE generation. The sensitivity is higher, but still not dramatic, for early and fast TQs, where the first injection contributes a relatively larger part of the total density increase. This trend is however only applicable to staggered injections; for single-stage injections, all pellets are doped, leading to a generally low impact of the drifts on both the RE generation and the material assimilation. Moreover, in such cases, an earlier and faster TQ prevents the shards from experiencing the highest temperatures present in the plasma, which would result in the longest drifts, so that the effect of the drift would be smaller rather than larger.

The sensitivity of the RE current to the drift dynamics may, however, be significantly higher when accounting for scrape-off RE losses,  as the plasma becomes vertically unstable and moves towards the wall \cite{KiramovBreizman, MartinSolis2022, Wang_2025, Bandaru_2025}. The proximity to the highly conducting wall means that the plasma will move approximately in such a way that the poloidal flux at the last closed flux surface remains constant \cite{Wang_2025}, which was recently exploited to develop an approximate model for scrape-off losses in DREAM \cite{Vallhagen_scrapeoff}. Thus, the formation of a large RE beam, aborting the drop in the poloidal flux, can significantly slow down the plasma motion and effectively eliminate the scrape-off losses during the RE generation phase. However, even a moderate reduction of the RE generation allows the poloidal flux to drop further and more flux surfaces to be scraped off, which further reduces the RE current, creating a positive feedback loop \cite{Vallhagen_scrapeoff, MartinSolis2022, Bandaru_2025}. 

This results in a rather sharp transition between cases forming a multi-MA RE beam and cases where all flux surfaces are scraped off before a macroscopic RE beam is formed. It was found in \cite{Vallhagen_scrapeoff} that this transition in ITER occurs for cases where the RE current without scrape-off losses reaches values in the $5-7\,\rm MA$-range. Thus, the relatively small changes in the RE current found in section~\ref{sec:ITER} could therefore still decide whether a large RE current is formed before all the flux surfaces are scraped off. This scenario is particularly plausible for the case shown in figure~\ref{fig:St4_bad_TQ}, with a fast and early TQ. The relevance of the first pellet in the staggered injection scheme, potentially including a small Ne doping to reduce the plasmoid drift, is therefore dependent on the vertical motion of the plasma. The combination of the effect of plasmoid drifts and scrape-off losses of REs thus appears to be a topic of interest for future studies.

\section{Conclusion}
\label{sec:conclusion}
The assimilation of pellets injected into a tokamak, for disruption mitigation as well as refueling, may be strongly affected by plasmoid drifts caused by the charge separation by the $\nabla B$-current and the subsequent $E\times B$-drift. We have implemented a semi-analytical model for plasmoid drifts in the numerical disruption modelling tool DREAM. The low computational cost of the plasmoid drift model enables simulating shattered pellet injections with hundreds of shards with the drift effect taken into account. 

This model was able to reproduce the density evolution following a shattered pellet injection in the AUG tokamak, with both pure D pellets and Ne-doped pellets, with values of the free model parameters that are consistent with previous measurements and simulations. Both the experimental measurements and our simulations corroborate the expectation that pure hydrogenic pellets are significantly more affected by plasmoid drifts than Ne-doped pellets. 
The predictive power of the model is, however, limited by its sensitivity to uncertainties in the free parameters, in particular the vertical half-width $\Delta y$ of the drifting cloud. Another source of uncertainty is the boundary condition for the drift motion, namely, what happens with the material drifring outside the last closed flux surface. 
However, despite the model uncertainties, the model could be used to estimate the effect of plasmoid drift in $15\,\rm MA$ H-mode ITER discharges with a disruption mitigated by a staggered shattered pellet injection. It was found that a first injection stage consisting of a pure hydrogenic pellet is likely to be strongly affected by plasmoid drifts, potentially rendering this injection stage ineffective. 
Our results also indicate that adding a small Ne-doping also to the first injection stage significantly increases the assimilation rate. This, however, requires that the Ne-doping is large enough to keep the plasmoid temperature from increasing over $\sim 5\,\rm eV$, but small enough to not significantly accelerate the onset of the thermal quench.

The generated runaway electron current was found to be largely insensitive to drift losses in the first injection stage, although some sensitivity was found when assuming a fast thermal quench with an early onset. The sensitivity could, however, increase when including scrape-off runaway losses, in which case the assimilation of the first pellet could decide whether a macroscopic runaway current is generated before all flux surfaces are scraped off. Therefore, it is necessary to take these losses into account when assessing the relevance of the first injection stage, and of attempting to suppress the plasmoid drift for this stage with a small Ne-doping.

\section*{Acknowledgement}
The authors are grateful to T. F\"ul\"op, I. Pusztai, A. Bock, M. Hoelzl, M. Kong, W. Tang, P. Lang, W. Zholobenko, S. Newton and E. Nardon for fruitful discussions. The work has been partly carried out within the framework of the EUROfusion Consortium, funded by the European Union via the Euratom Research and Training Programme (Grant Agreement No 101052200 — EUROfusion). Views and opinions expressed are however those of the authors only and do not necessarily reflect those of the European Union or the European Commission. Neither the European Union nor the European Commission can be held responsible for them. The work was also supported by the Swedish Research Council (Dnr.~2022-02862) and the Adlerbert Research Foundation.
The ASDEX Upgrade SPI project has been implemented as part of the ITER DMS Task Force programme. The ASDEX Upgrade SPI system and related diagnostics have received funding from the ITER Organization under contracts IO/20/CT/43-2084, IO/20/CT/43-2115, IO/20/CT/43-2116.

\bibliography{bibliography.bib}
\appendix
\section{Scaling of the drift with the cloud width}
\label{app:Delta_y}
In the limit of a high pressure cloud, where the pressure equilibration time is long compared to the drift duration, one may derive a power law scaling of the distance drifted $\Delta R$ with the cloud half-width $\Delta y$. For this purpose, the distance drifted can be estimated by $\Delta R \sim v_\mathrm{drift} t_\mathrm{drift}$, where $v_\mathrm{drift}$ is a representative drift speed and $t_\mathrm{drift}$ is the duration of the drift. In the limit considered here, $t_\mathrm{drift}$ is proportional to the time $t_\mathrm{pol}$ it takes the cloud to expand one poloidal turn around the plasma. At this point, the $\nabla B$-current in the inboard and outboard side of the cloud cancel each other, analogously to a tokamak equilibrium. This time only scales with the size of the tokamak and the cloud expansion speed, which is essentially equal to the speed of sound inside the cloud. Thus, the drift duration does not scale with $\Delta y$, assuming the cloud temperature is independent of $\Delta y$, so that the scaling with $\Delta y$ only enter via $v_\mathrm{drift}$. The scaling of $v_\mathrm{drift}$ with $\Delta y$ can be obtained from the following considerations:

\begin{itemize}
\item The equation of motion for the cloud has the form (see equation 2.24 in \cite{Vallhagen2023Drift}) 
\begin{equation}
    \bar{n} \dot{v} = F - \gamma v, 
    \label{eq:drift_forces}
\end{equation}
where $\bar{n}$ is the line integrated density, $F$ is the driving force density behind the drift and $\gamma$ represents the friction associated with the currents leaving the cloud parallel to the field lines. The scaling of the representative drift velocity may be estimated by taking the quasi-steady state solution of equation \ref{eq:drift_forces}, $v_\mathrm{drift}\sim F/\gamma$.

\item The driving force is proportional to the excess vertical $\nabla B$-current, not compensated for by the $\nabla B$-current in the background plasma. In case of a continuous flow of the ablated material, as assumed in \cite{Vallhagen2023Drift}, the vertical surface area of the cloud does not scale with $\Delta y$, so that the scaling of $F$ with $\Delta y$ comes from that $F\sim p$, where $p\sim \bar{n}T/L$ is the cloud pressure, $T$ is the cloud temperature and $L$ is the length of the cloud. Assuming a constant cloud temperature, and thus a constant sound speed, both $T$ and $L$ are independent of $\Delta y$, i.e. $F\sim \bar{n}$. 

\item The pellet will spread the ablated material within a radius similar to $\Delta y$, providing a line integrated density source $\dot{\bar{n}}\sim \dot{N}/\Delta y^2$, where $\dot{N}$ is the ablation rate, for the time $t_\mathrm{det}$ it takes for the cloud to detach from the pellet. Since both $F$ and the left hand side in equation \ref{eq:drift_forces} are proportional to $\bar{n}$, the initial acceleration (neglecting the friction term since $v$ is small before the detachment) is constant and independent of $\bar{n}$, and hence also of $\Delta y$. The detachment time is therefore given by the time it takes the cloud to drift a distance $\Delta y$ with a constant acceleration $\dot{v}_0$, i.e. 
\begin{equation}
    \Delta y = v_\mathrm{p} t_\mathrm{det} + \frac{\dot{v}_0t_\mathrm{det}^2}{2}, 
    \label{eq:drift_acc}
\end{equation}
where $v_\mathrm{p}$ is the pellet speed. Thus $t_\mathrm{det}\sim \Delta y$ or $t_\mathrm{det}\sim \sqrt{\Delta y}$, depending on if the detachment time is dominated by the pellet speed or the drift acceleration. This gives
\begin{equation}
    F\sim \bar{n} \sim \frac{t_\mathrm{det}}{\Delta y^2} \sim \Delta y ^{-1} \quad \mathrm{or} \quad \Delta y ^{-1.5}.
\end{equation}

\item The friction force is proportional to the current leaving the cloud parallel to the field lines, which is dominated by the Ohmic current for background plasma temperatures $T_\mathrm{bg}\gtrsim 100 \,\rm eV$. The Ohmic current scales as $I_\Omega\sim \rho A U/L_\mathrm{con}$, where $\rho$ is the plasma resistivity, $A =  \Delta y \delta R$ is the poloidal cross-section area of the portion of the cloud considered, with an extent $\delta R$ in the major radius direction. Moreover, the potential difference between the upper and lower part of the cloud scales as $U \sim E_y\Delta y$, where $E_y\sim v_\mathrm{drift}B$ is the electric field associated with the $E\times B$ drift speed. The typical length $L_\mathrm{con}$ of the field lines connecting the end caps of the cloud is inversely proportional to the probability $p_\mathrm{con}\sim \Delta y/r$ that a field line will connect to the cloud after one turn around the torus, i.e. $L_\mathrm{con}\sim R/p_\mathrm{con} = Rr/(\Delta y)$. In combination,
\begin{equation}
    \gamma \sim \frac{I_\Omega}{v_\mathrm{drift}}\sim \Delta y ^3.
\end{equation}
\end{itemize}
Finally, this gives
\begin{equation}
    \Delta R \sim \frac{F}{\gamma} \sim \Delta y^{-4} \quad \mathrm{or} \quad \Delta R \sim \Delta y^{-4.5}
    \label{eq:Delta_y_scaling}
\end{equation}
depending on which term dominates on the right hand side of equation \ref{eq:drift_acc}.

To summarise, the scaling of $\Delta R$ with $\Delta y$ comes from: the $\Delta y ^{-2}$ scaling of the line integrated ablation source; the $\Delta y^1$ or $\Delta y^{0.5}$ scaling of the detachment time; the $\Delta y^1$ scalings of the poloidal corss section area and potential difference between the upper and lower part of the cloud; and the $\Delta y ^{-1}$ scaling of the probability that a field line will reconnect to the cloud after one turn around the torus. Here we have considered a continuous flow of ablated material, rather than a striated flow divided into discrete plasmoids, but this does not affect the scaling; for a cylindrical plasmoid, with a radius $\Delta y$, the driving force $F$ would have another factor $\Delta y$ from the vertical surface area. However, the poloidal cross-section area $A$ would be proportional to $\Delta y^2$, and these additional factors cancel each other. Note, however, that we have assumed that e.g. the cloud temperature and expansion speed are independent of $\Delta y$. The scalings in equation \ref{eq:Delta_y_scaling} could therefore be altered by such interdependencies.
\end{document}